\documentclass[prb,superscriptaddress,twocolumn,showpacs]{revtex4}
\usepackage{amsmath}
\usepackage{graphicx}
\usepackage{epsfig}
\begin{document}
\title{Stark effect on the exciton spectra of vertically coupled quantum dots: horizontal field orientation and
non-aligned dots}
\author{B. Szafran} \affiliation{Faculty of Physics and Applied
Computer Science, AGH University of Science and Technology, al.
Mickiewicza 30, 30-059 Krak\'ow, Poland}

\author{F.M. Peeters}
\affiliation{Departement Fysica, Universiteit Antwerpen,
Groenenborgerlaan 171, B-2020 Antwerpen, Belgium}
\author{S. Bednarek} \affiliation{Faculty of Physics and Applied
Computer Science, AGH University of Science and Technology, al.
Mickiewicza 30, 30-059 Krak\'ow, Poland}
\date{\today}

\begin{abstract}
We study the effect of an electric-field on an electron-hole pair in
an asymmetric system of vertically coupled self-assembled quantum
dots taking into account their non-perfect alignment. We show that
the non-perfect alignment does not qualitatively influence the
exciton Stark effect for the electric field applied in the growth
direction, but can be detected by application of a perpendicular
electric field. We demonstrate that the direction of the shift
between the axes of non-aligned dots can be detected by rotation of
a weak electric field within the plane of confinement. Already for a
nearly perfect alignment the two-lowest energy bright exciton states
possess antilocked extrema as function of the orientation angle of
the horizontal field which appear when the field is parallel to the
direction of the shift between the dot centers.
\end{abstract}
 \maketitle
\section{Introduction}

The effect of the electric field on the exciton recombination energy
is used to probe the properties of the electron and hole confinement
in separate self-assembled quantum dots \cite{Fry,TCK}  through a
strong deformation of the carrier wave functions. For vertically
coupled quantum dots\cite{xie} the electric field oriented in the
growth direction leads to a redistribution of the carriers between
the dots.\cite{ek1,ek2,Krenner2,ek3,chwiej} Typical electric field
applied for vertically coupled quantum dots is of the order of 20-30
kV/cm, i.e., by an order of magnitude smaller than the one used to
probe the confinement potential\cite{Fry} of a single dot.
Therefore, such experiments on the electric-field effect for coupled
dots resolve rather the properties of the molecular coupling than
the fabrication dependent details of the confinement in separate
dots. Moreover, the charge redistribution induced by the electric
field between the different dots have naturally a much stronger
energy effect than deformation of the wave functions inside each of
the dots. Due to these features, previous simple
modeling\cite{chwiej} (done in parallel with the experimental
work\cite{ek1}) using a square well vertical and harmonic lateral
confinement\cite{baer} led to the correct description of the
observed \cite{ek1,ek2} electric field induced dissociation of the
electron-hole pair in an avoided crossing of dark and bright energy
levels.\cite{stare} It also successfully predicted\cite{chwiej}
non-trivial features of the photoluminescence spectrum observed
during the negative trion dissociation by the electric
field.\cite{Krenner2}

A vertical electric field was used very recently\cite{ek3} to verify
the growth process of intentionally strongly asymmetric double
quantum dots. In this paper we demonstrate that the horizonal
electric field (perpendicular to the growth direction) can be useful
to estimate the non-perfect vertical alignment of the dots and to
determine the direction of the horizontal shift between them. To
account for the horizontal field effects we replaced the harmonic
lateral profile of our previous model\cite{chwiej} by a quantum well
and developed a computational approach for treating excitons in
systems with no axial symmetry and for an arbitrary electric field
direction.

The vertical electric field-induced redistribution of charge
carriers is smooth only for the electron which tunnels much more
effectively between the dots and is responsible for the most
characteristic experimental features of the spectrum.
\cite{ek1,ek2,Krenner2,ek3,chwiej} Similarly, it is the electron
redistribution between the dots due to a rotation of a weak
horizontal field that allows for the detection of the non-perfect
alignment, the hole charge being only shifted within the separate
dots. Therefore, we decided to use for simplicity the single band
model to describe the hole.

The paper is organized as follows, Section II describes the
computational approach, our numerical results are given in Section
III, and Section IV contains the summary and conclusions.
\begin{figure}[htbp]
  \hbox{\epsfysize=65mm
                \epsfbox[16 600 300 840] {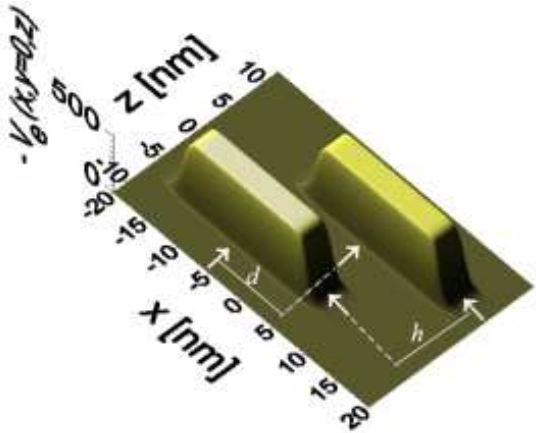}\hfill}
\caption{The electron confinement potential [Eq. (2)] in meV, in the
$y=0$ plane, for the offset between the axes of the dots $d=10$ nm
(equal to the dot radius $R$) and the vertical distance between the
dot centers of $h=10$ nm. The height of both the dots is $2Z=4$ nm.
The lower (negative $z$) dot has depth $V_l^e=518$ meV, the dot for
positive $z$ has depth of $V_u^e=508$ meV.} \label{pot}
\end{figure}

\section{Double dot model and method of calculations}
We work in the single valence band approximation, for which the
Hamiltonian of the electron-hole pair is
\begin{eqnarray}
H&=&-\frac{\hbar^2}{2m_e}\nabla_e^2-\frac{\hbar^2}{2m_h}\nabla_h^2+V_e({\bf
r}_e)+V_h({\bf r}_h)\nonumber \\ &&-\frac{e^2}{4\pi \epsilon
\epsilon_0 |{\bf r}_e-{\bf r}_h|}-e{\bf F} \cdot \left({\bf r}_e
-{\bf r}_h\right),
\end{eqnarray}
with ${\bf r}_e$, ${\bf r}_h$ - electron and hole coordinates,
$m_e$, $m_h$ - effective band masses for the electron and the hole,
$\epsilon$ the dielectric constant, and ${\bf F}$ is the electric
field. The electron confinement potential of the double dot system
is taken in the form of a sum of double disk-shaped quantum wells
with smooth walls (see Fig. \ref{pot})
\begin{eqnarray}
V_e({\bf r})=&& \\ & -V_e^l/
(1+\left(\frac{(x-d/2)^2+y^2}{R^2}\right)^{10})
(1+\left(\frac{(z-h/2)^2}{Z^2}\right)^{10})\nonumber \\
& - V_e^u/
(1+\left(\frac{(x+d/2)^2+y^2}{R^2}\right)^{10})(1+\left(\frac{(z+h/2)^2}{Z^2}\right)^{10})
\nonumber,
\end{eqnarray}
here $R$ stands for the radius of a single dot, $Z$ is half of its
height, the dots centers are placed symmetrically with respect to
the origin at points $(x,y,z)=\pm (d/2,0,h/2)$ (the spacer layer is
then equal to $b=h-2Z$), $V_e^l$ and $V_e^u$  are the lower and
upper dot depths, respectively. For simplicity we take the same size
of both dots, the inevitable asymmetry between them (due to size
difference, indium composition, nonsymmetric strain, etc.) can for
the limited purpose of the present paper be effectively taken into
account by applying different depths of the dots. The confinement
potential for the hole is of the same form, only with depths
$V_h^l$, $V_h^u$ replacing $V_e^l$ and $V_e^u$. The bottom of the
conduction band of the barrier material (GaAs) is taken as the
reference energy for electrons, and the top of the GaAs valence band
as the reference level for the holes, i.e., the eigenvalues of (1)
have to be shifted up by the GaAs band gap to give the photon energy
measured in a luminescence experiment.

\begin{figure*}[htbp]
  \hbox{\epsfysize=65mm
                \epsfbox[16 600 600 840] {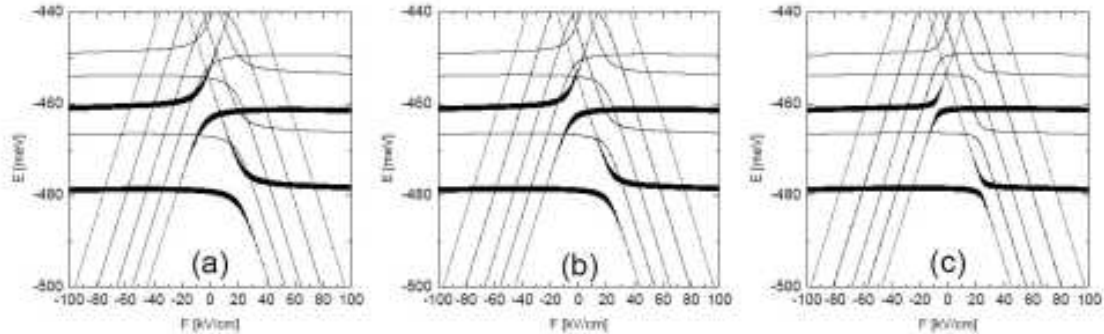}\hfill}
\caption{ \label{h10pz} The exciton spectra for the electric field
oriented along the $z$ direction for a vertical distance between the
dots centers $h=10$ nm for (a) perfectly aligned dots $d=0$, (b)
horizontal distance between the vertical symmetry axis of the dots
$d=5$ nm, and $d=10$ nm (c). }
\end{figure*}

\begin{figure}[htbp]
  \hbox{\epsfysize=65mm
                \epsfbox[16 130 530 840] {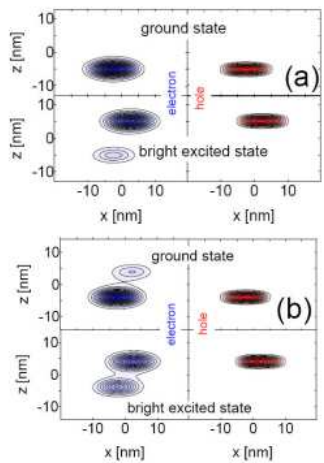}\hfill}
\caption{ \label{510} The electron and hole probability density
distributions for the ground state and the first excited bright
state in the absence of the electric field for $d=5$ nm and $h=10$
nm (a), and $h=8$ nm (b).}
\end{figure}

We previously\cite{chwiej} demonstrated, that for an interacting
electron-hole pair both the energy spectrum and the particle
distribution are nearly the same when the asymmetry of the double
well is exclusively introduced for one of the particles, since it is
translated by the interaction into an asymmetry for the other
particle. In this paper we assume that the dot situated on the
negative side of the $z=0$ plane is deeper by 10 meV for both the
electron and the hole $V_e^l=V_e^u+10$ meV, $V_h^l=V_h^u+10$ meV,
and $V_e^u=508$ meV, $V_h^u=218$  meV is taken the same as in our
previous paper\cite{chwiej} assuming that the dots embedded in a
pure GaAs matrix are made of Ga$_{x}$In$_{1-x}$As alloy with
$x=0.66$. Accordingly\cite{chwiej} we take $m_e=0.037$, $m_h=0.45$,
and $\epsilon=12.5$. The size parameters are taken from
Ref.\cite{Krenner2}, diameter $2R=20$ nm and height $2Z=4$ nm
(therefore $b=h-4$ nm).

The Schroedinger equation with Hamiltonian (1) is solved using the
exact diagonalization (configuration interaction) approach keeping a
complete account of the electron-hole correlation. As the basis for
the electron-hole pair we take products of single-electron states
and single-hole states
\begin{equation}
\phi({\bf r}_e,{\bf r}_h)=\sum_{j,k} d_{jk} f_e^{(j)}({\bf
r}_e)f_h^{(k)}({\bf r}_h),
\end{equation}
where $f_e^{(j)}$, $f_h^{(k)}$ are the single-electron and
single-hole wave functions of state $j$ and $k$, respectively.
 The single-particle wave functions are obtained in the
multicenter basis of displaced Gaussian functions,
\begin{eqnarray}
f (r)&=& \sum_{i}^M c_i \exp\left[-\alpha_i
\left((x-x_i)^2+(y-y_i)^2\right)\nonumber \right. \\ &&\left.
-\beta_i (z-z_i)^2\right],
\end{eqnarray}
where $c_i$ are the linear variational parameters, $\alpha_i$ and
$\beta_i$ are the nonlinear variational parameters describing the
strength of $i-th$ Gaussian around point $(x_i,y_i,z_i)$, whose
coordinates are variational parameters by themselves. We took 11
basis functions for each of the dots: one situated at the dot center
$\pm(d/2,0,h/2)$, eight around it within the plane of confinement
$z_i=\pm h/2$, on a circle with optimized diameter. Two additional
centers are placed above and below the center of each dot
$\pm(d/2,0,h/2+\pm \Delta)$ with $\Delta$ optimized variationally.
Addition of more centers outside the plane of confinement or
including more centers within it does not improve the results
significantly. In the configuration basis (3) we include all the
single-particle wave functions obtained in a diagonalization of the
electron and the hole Hamiltonians each in a basis of form (4) (with
position of centers and other nonlinear parameters optimized
separately for the electron and hole), which eventually yields 484
{\it localized} basis functions to treat the exciton. The
recombination probability for the exciton state with wave function
$\phi$ is calculated from the envelope wave function as
\begin{equation} p=\left |\int d^6{\bf r} \phi({\bf
r}_e,{\bf r}_h)\delta^3({\bf r}_e-{\bf r}_h) \right |^2 .
\end{equation}

Two-dimensional version of the multicenter basis was previously used
for laterally coupled dots.\cite{Bsz} Our study\cite{asy} on
electron systems in a single circular dot performed with a similar
technique showed that the superposition of Gaussians very well
approximates the angular momentum eigenstates.
\section{Results}

\begin{figure*}[htbp]
  \hbox{\epsfxsize=160mm
                \epsfbox[16 650 580 840] {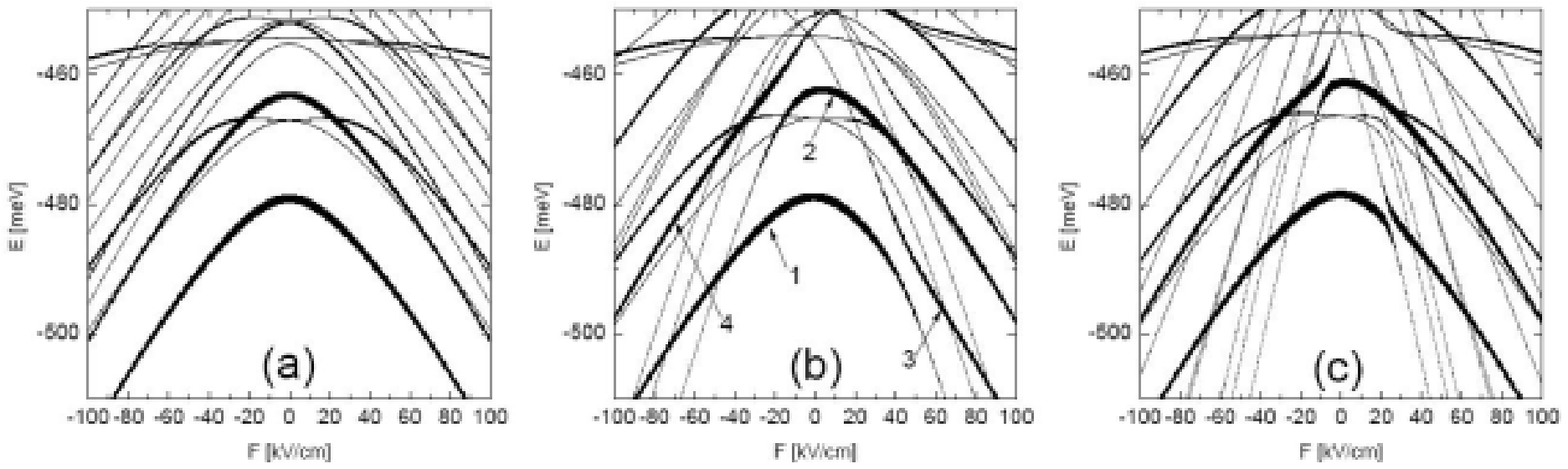}\hfill}

\caption{ \label{h_10px} The exciton spectra for the electric field
 oriented along the $x$ direction for the vertical distance
between the dots centers $h=10$ nm for (a) perfectly aligned dots
$d=0$, (b) horizontal distance between the vertical symmetry axis of
the dots $d=5$ nm, and (c) $d=10$ nm. Single-particle densities for
the bright energy levels labeled by numbers 1-4 in (b) are presented
in Fig. 6.}
\end{figure*}
\begin{figure}[htbp]
  \hbox{\epsfysize=65mm
                \epsfbox[16 130 530 840] {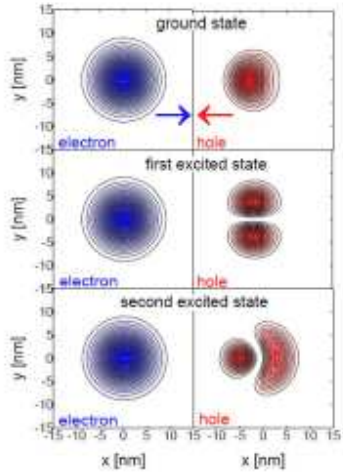}\hfill}
\caption{ \label{010gora} Electron and hole densities for 3
lowest-energy states for $d$=0 nm and $h=$ 10 nm taken within the
plane of confinement of the lower, deeper dot ($z=-5$ nm). For the
field of $F=20$ kV/cm parallel to the $x$ axis (arrows at the
electron and hole panels at the top-left side of the figure show the
direction of the electric force acting on the electron and the
hole). The energy levels of these states are displayed in Fig.
\ref{h_10px}(a).}
\end{figure}

The effect of the unperfect alignment of the dots for the exciton
Stark effect due to the electric field oriented in the growth
direction is presented in Fig. \ref{h10pz}. The figure shows the
dependence of the exciton spectra on the vertically oriented
electric field for the barrier thickness $b=6$ nm ($h=10$ nm).  For
the dots aligned vertically (a) as well as for the offset between
the axes of $d=5$ nm (b) and $d=10$ nm (c) at $F=0$ we observe two
bright lines (thickness of lines is set proportional to the
recombination probability). The particle distribution for the bright
states is displayed in Fig. \ref{510}. Fig. \ref{510}(a) shows the
cross section ($y=0$) of the electron and hole probability density
distributions (obtained through integration of the two-particle
probability density in the coordinates of the other particle) for
both the bright states at $F=0$. We notice that in these states the
hole is entirely localized in one of the dots, and the electron
tends to accompany the hole. In the ground-state the electron is
entirely localized in the deeper (lower) dot. In the excited bright
state, the electron is predominantly localized in the upper dot, but
a small part of the electron wave function leaks to the other
(deeper) dot. For thinner spacer layer $b=4$ nm ($h=8$ nm) a leakage
of the electron density to the dot without the hole appears in the
ground state [Fig. \ref{510}(b)], and in the excited state it is
strengthened. Note, that in the excited state the hole density
maximum is displaced to the left within the upper dot following the
electron density noticeably shifted to the lower dot.

 The
vertically oriented electric field $F>0$ tends to push the electron
to the upper (shallower) dot. In the ground state near $F=20$ kV/cm
we see [cf. Fig. \ref{h10pz}(a-c)] an avoided crossing of the bright
energy level with both carriers in the deeper dot with a dark energy
level of separated carriers (hole stays in the lower dot, the
electron is transferred to the upper one). The crossing is avoided
due to the electron tunnel coupling between the dots. In the upper
bright state - no avoided crossing is observed for $F>0$, the
positive electric field stabilizes the electron in the upper dot. On
the other hand, for the upper bright energy level, the negative
electric field of smaller absolute value (around 10 kV/cm) is enough
to transfer the electron from the upper to the lower (deeper) dot
which results in an avoided level crossing similar to the one
appearing in the ground-state [see the upper avoided crossing at
left of $F=0$ in Fig. \ref{h10pz}(a-c)].

We see that for the displaced axes of the dots [Figs. \ref{h10pz}
(b-c)] the dependence of the spectra on the vertical electric field
is qualitatively similar to the perfect alignment case [Fig.
\ref{h10pz}(a)] but with an avoided level crossings width which is
smaller due to the suppressed electron tunnel coupling as the
barrier thickness in increased.

\begin{figure*}[htbp]
  \hbox{\epsfxsize=150mm
                \epsfbox[16 650 580 840] {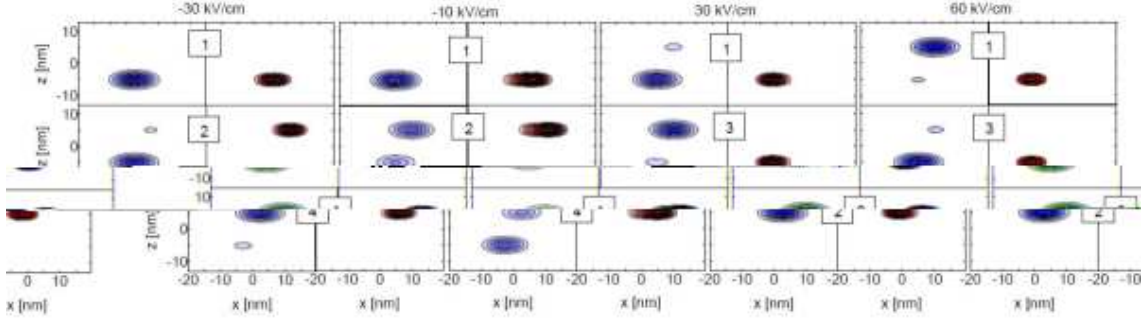}\hfill}

\caption{ \label{5_10fp} The $y=0$ cross section of the electron and
hole probability density distributions for the ground state and the
first excited bright state for different values of the electric
field in the $x$-direction (given at the top of the figure for each
set of plots) for $d=5$ and $h=10$ nm. The selected bright states
denoted by numbers 1-4, as in Fig. \ref{h_10px}(b). The contour
plots for states 1 and 2 at $F=0$ were shown in Fig. \ref{510}.}
\end{figure*}

\begin{figure}[htbp]
 \hbox{\epsfxsize=60mm
                \epsfbox[0 0 550 820] {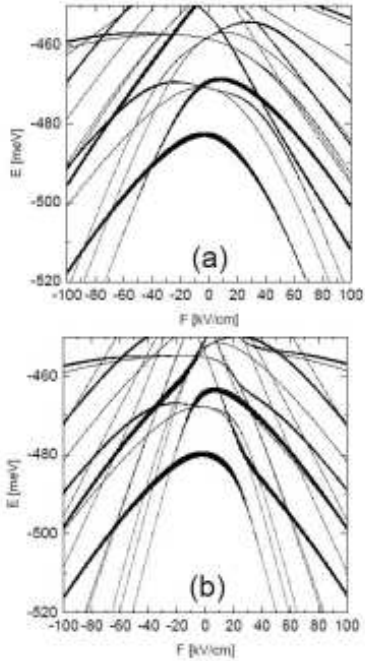}\hfill}
\caption{ \label{h8px} The exciton spectra for the electric field
oriented along the $x$ direction for a vertical distance between the
dots centers $h=8$ nm for (a) horizontal distance between the
vertical symmetry axis of the dots $d=5$ nm. (c) and $d=10$ nm. }
\end{figure}

\begin{figure}[htbp]
 \hbox{\epsfxsize=60mm
                \epsfbox[0 0 550 820] {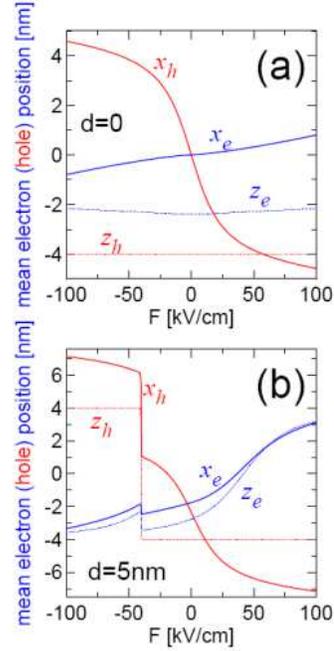}\hfill}
\caption{ \label{positions} The ground-state mean electron and hole
position dependence on the horizontal electric field oriented along
the $x$ direction for $h=8$ nm (a) perfectly aligned dots and (b)
offset of the dots axes of $d=5$ nm. The energy levels for
parameters used in (b) are displayed in Fig. \ref{h8px}(a).}
\end{figure}

\begin{figure*}[htbp]
 \hbox{\epsfxsize=120mm
                \epsfbox[0 460 550 834] {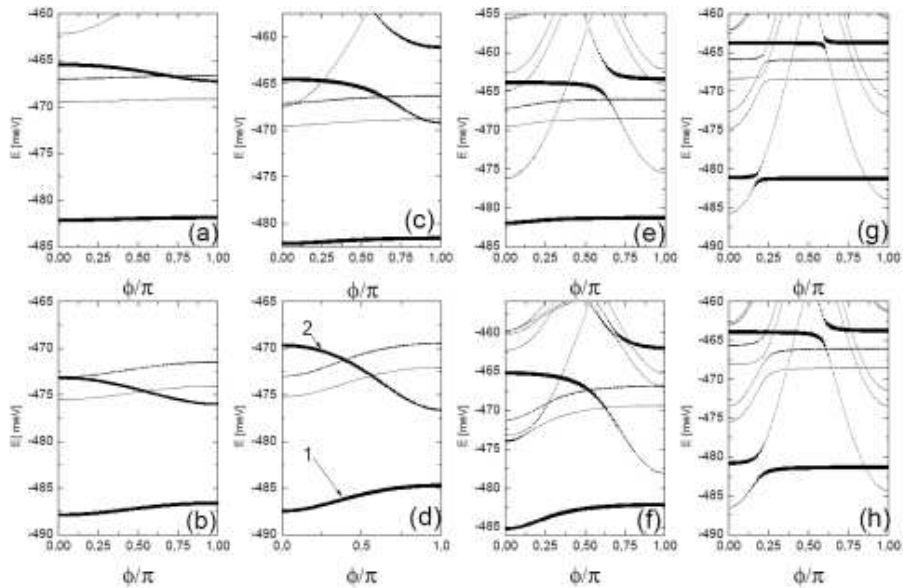}\hfill}
\caption{ \label{obroty} The exciton spectra for the electric field
$F=20$ kV/cm oriented in the $x-y$ plane as function of the angle
between the field and the $x$ axis (${\bf
F}=F(\cos(\phi),\sin(\phi),0)$). The plots in the upper (lower) row
correspond to $h=10$ nm ($h=8$ nm). Figures (a-b), (c-d), (e-f) and
(g-h) correspond to $d=2,5,10$ and 15 nm, respectively. Electron and
hole densities for energy levels of plot (d) are displayed in Fig.
\ref{ffobroty}.}
\end{figure*}

\begin{figure}[htbp]
 \hbox{\epsfxsize=60mm
                \epsfbox[0 200 560 834] {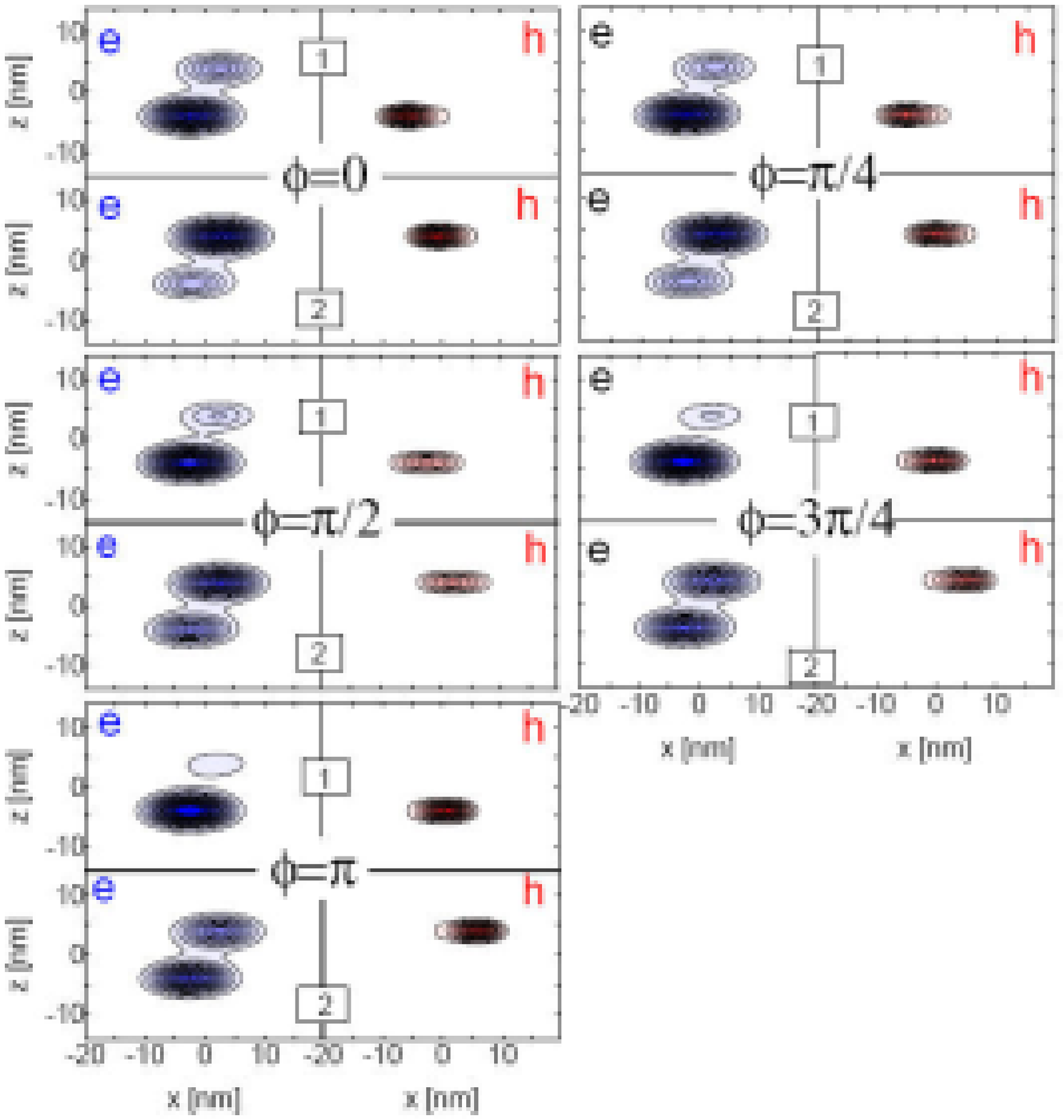}\hfill}

\caption{ \label{ffobroty} The $y=0$ cross section of the electron
(the panels marked by 'e') and hole (the panels marked by 'h')
probability density distributions for the ground state (denoted by 1
in a square frame) and the first excited bright state (denoted by 2)
for $d=5$ and $h=8$ nm. The corresponding energy levels are
displayed and marked by the same numbers in Fig. \ref{obroty}(d).
The horizontal axis corresponds to $x$ variable, the vertical - to
$z$ variable. The electric field is kept at $|{\bf F}|$=20 kV/cm,
each panel corresponds to a different orientation of the electric
field within the $x-y$ plane given by $\phi$ which is the angle
between the {\bf F} and the $x$ axis. The left plots in each column
show the electron distribution and the right ones--the hole
distribution The plots for states 1 and 2 at $F=0$ were shown in
Fig. \ref{510}(b) (ground and excited state,
respectively).}\end{figure}

Pronounced qualitative differences of the spectra for aligned and
not aligned dots appear in case the electric field is oriented
horizontally. Fig. \ref{h_10px}(a) shows the spectrum for perfectly
aligned dots.  For both the bright states (with both carriers
localized in the same dot) the electric field separates the electron
and the hole within each dot pushing the carriers at its opposite
sides, which leads to a decreased recombination probability and a
decreased energy. The first excited state, which is dark and twice
degenerated at $F=0$, corresponds to both carriers in the deeper
dot, but with a hole excitation (of $p$ type). The horizontal
electric field lifts the degeneracy of the $p$ energy level of the
hole.  The electron (left panel) and the hole (right panel) density
for $F=20$ kV/cm applied in the $x$ direction is shown in Fig.
\ref{010gora} for the ground-state and the two lowest-energy excited
states with $p$-excitations of the hole. The plot was made in $x,y$
coordinates for the plane of confinement of the lower dot $z=-h/2$
[all the three states correspond to carriers totally localized in
the lower dot cf. upper panels of Fig. \ref{510}(a)]. The lower dark
energy state with $p$-hole excitation has a nodal surface at the
plane $y=0$, and the parity with respect to this plane is conserved
when the field is applied in the $x$-direction (see the panel of
Fig. \ref{010gora} for the first excited state), hence the zero
recombination probability for all $F$. The other $p$-level, higher
in energy becomes bright at larger $F$, when the hole parity with
respect to the direction perpendicular to the field is destroyed
(see the panel of Fig. \ref{010gora} for the second excited state).
We note that in all the three states, for $F=20$ kV/cm the electron
density is not perturbed by the horizontally applied field, but its
effect on the hole (for which the confinement is weaker) is clearly
visible.

For non-perfectly aligned dots the component of the electric field
parallel to the direction of their relative horizontal displacement
redistributes the carriers between the dots. The low-energy spectrum
plot for $d=5$ nm ($h$ still equal to 10 nm) contains 4 bright
energy levels denoted by 1, 2, 3, and 4 in Fig. \ref{h_10px}(b). The
modification of the carrier distribution by the electric field in
these states is shown in Fig. \ref{5_10fp}. In the ground state [for
$F=0$ both carriers in the lower dot - see Fig. \ref{510}(a)] the
electron passes to the right (shallower) dot near $F=+60$ kV/cm. On
the other hand, in the excited state which is bright at $F=0$ and
marked by '2' in Fig. \ref{h_10px} both carriers are localized in
the upper shallower dot. The electron is transferred to the left dot
(lower and deeper) for the field around $F=-30$ kV/cm. Both these
electron transfers (ground state at +60 kV/cm and excited bright
state at -30 kV/cm) are associated with avoided crossings of bright
and dark energy levels. These avoided crossings become smaller for
an offset of 10 nm between the axes of the dots, the case presented
in Fig. \ref{h_10px}(c).

Fig. \ref{h8px} shows the spectra for stronger interdot coupling
($h=8$ nm). Compared to $h=10$ nm case, the weak-field extrema of
the energy levels are shifted more distinctly off the $F=0$ point
due to a larger value of the built-in dipole moment resulting from
the shift of the electron density related to the tunnel coupling
[see Fig. \ref{510}(b) and notice opposite shifts for the extrema of
two lowest bright energy levels and much larger value of the dipole
moment for the excited state].  The dipole moment
($\mu=|e|\left(<{\bf r}_h>-<{\bf r}_e>\right)$ is proportional to
the difference in the mean positions of the electron and the hole.
The impact of the horizontal field ($x$) on the spatial position of
particles in the ground-state is plotted for $h=8$ nm in Fig.
\ref{positions}. For perfect alignment $d=0$ [Fig.
\ref{positions}(a)] the field applied in $x$ direction does not
affect the $z$ position of the particles, although one can notice a
shallow minimum for the electron at $F=0$. The horizontal field
slightly strengthens the electron confinement within each of the
dots thus enhancing the interdot vertical coupling and weakly
shifting the electron towards the upper dot. Fig. \ref{positions}(a)
shows a stronger reaction of the horizontal position of the hole
than of the electron, which was already noted in the context of Fig.
\ref{010gora}. For non-perfect alignment [$d=5$ nm, see Fig.
\ref{positions}(b)] the horizontal field ($x$) leads to a very
strong dependence of the vertical positions ($z$) of the particles.
For $F>0$ the electron is transferred to the dot at right and is
shifted to the top while hole is stabilized in the left dot. Due to
the electron tunnel coupling the electron position is continuous and
a smooth function of $F$. On the contrary the ground-state values of
the hole position have a strong jump at $F<-40$ kV/cm, where the
crossing of energy levels appear in the ground-state  [cf. Fig.
\ref{h8px}(a)]. Note that also the electron positions are modified
during this jump - the electron, localized in the lower (left) dot
tends to follow the hole when it leaves to the upper (right) one.

Fig. \ref{obroty} shows the low-energy exciton spectrum for a fixed,
rather small, length of the electric field of $F=20$ kV/cm as
function of the angle $\phi$ that it forms with the $x$ axis, ${\bf
F}=F(\cos(\phi),\sin(\phi),0)$. The upper panel of the figure
corresponds to $h=10$ nm and the lower to $h=8$ nm. The plots from
left to right are calculated for offsets $d=2,5,10$ and 15 nm (for
perfectly aligned dots $d=0$ the spectra are independent of $\phi$).
Already for the smallest offset [Figs. \ref{obroty} (a-b)] we notice
that the ground state energy is minimal when the field is oriented
to the right ($\phi=0$), i.e., when it tends to transfer the
electron to the shallower right dot, and maximal when the field is
oriented to the left (tends to keep the electron in the deeper --
left dot). For the excited bright state (when both carriers tend to
remain in the shallower -- right dot), the energy is maximal when
the field pushes the electron to the right (reduced penetration of
the electron to the other - deeper dot) and minimal in the opposite
case (enhanced tunneling to the other dot). The energy dependence of
the excited bright state on $\phi$ is stronger than for the ground
state. This is due to the larger electron tunnel coupling for the
excited state [see Fig. \ref{510}]. The electron is simply more
willing to pass to the deeper dot.

For a larger offset of $d=5$ nm [Fig. \ref{obroty}(c-d)] the
dependencies of the bright energy levels on the angle $\phi$
preserve their character but become more pronounced. The electron
and hole densities in the $y=0$ plane are plotted in Fig.
\ref{ffobroty} for the two lowest-energy bright energy levels of
Fig. \ref{obroty}(d) and for opposite field orientation. They are to
be compared to the $F=0$ case displayed in Fig. \ref{510}(b), which
shows that in the absence of the field the probability to find the
electron in the upper dot for the ground state was smaller than the
probability to find the electron in the lower one in the excited
bright state. Field of 20 kV/cm at $\phi=0$ ($F$ oriented in $x$
direction) reverses this relation, as the field shifts more of the
electron to the upper dot when in the ground state and removes the
electron from the lower dot when in the excited state. A shift of
the hole distributions to the left and the squeeze of the
distribution to the left side of the dot is also visible. For the
electric field oriented along the $y$ direction ($\phi=\pi/2$) the
electron distribution of the carriers between the dots is similar to
the $F=0$ case. For the field oriented antiparallel to the $x$ axis
($\phi=\pi$) the electron is almost entirely localized in the left
dot when in the ground state and almost completely so when in
excited state.

The anti-locking of the energy extrema as function of $\phi$ for the
two lowest bright states observed in Figs. \ref{obroty} (a-d) at the
electric field orientation matching the direction of the shift
between the dots can be explained in the following way. The electric
field leads to a decrease of the energy levels through the
separation of the electron and hole charges [see Eq. (1)]. In the
ground state [see Fig. \ref{ffobroty}] the field at the angle
$\phi=0$ ($+x$ direction) enhances the separation stimulating the
electron to leave the hole [hence the energy minimum see Fig. 9(d)],
and at $\phi=\pi$  ($-x$ direction) prevents the electron-hole
separation (hence the maximum). In the excited bright state the
effect of the field on the carrier separation is opposite. For the
excited state the modulation of the recombination probability is
more pronounced, the electron coupling and the electron switching
between the dots is stronger, hence the more pronounced energy
dependence on $\phi$ [Fig. 9(d)].

For the offset of dot axes equal to the radii $d=10$ nm -- Figs.
\ref{obroty}(e-f) the excited bright energy level is involved in an
avoided crossing with a dark energy level. For even larger offset
$d=15$ nm -- Figs. \ref{obroty}(g-h) an avoided crossing appears
also in the ground state. The appearance of these anticrossings is
due to the stronger energetic effect of constant $F$ for increased
$d$.

\section{Summary and conclusions}
In summary, we have shown that for vertically coupled dots their
non-perfect alignment does not qualitatively influence the exciton
Stark effect of the electric field oriented in the growth direction.
 Although for
perfectly aligned dots the horizontal electric field only deforms
the electron and hole density within each of the dots, for
non-perfect alignment it leads to a redistribution of the particles
between the dots, i.e., in the direction perpendicular to the field.
We demonstrated that due to the relatively strong electron tunnel
coupling and strong electron confinement within each of the dots the
horizontal electric field can be used to tune the electron
distribution in the vertical direction only weakly affecting the
electron charge distribution within each of the dots.  On the other
hand, due to a weaker hole confinement the horizontal field
distinctly affects the hole distribution within each of the dots,
and the hole transfer between the dots is rapid due to negligible
tunnel coupling. We have shown that a rotation of the electric field
vector within the plane of confinement can be used to detect the
non-perfect alignment. To determine the direction of the horizontal
shift between the non-aligned dots, and to estimate the length of
the shift, they can be observed from the exciton photoluminescence
spectrum.

{\bf Acknowledgments} This work was supported by the EU network of
excellence SANDiE, the Belgian Science Policy and the Foundation for
Polish Science (FNP).

\end{document}